\documentclass[11pt,onecolumn]{scrartcl}
\usepackage{float,afterpage,longtable,amssymb,amsmath,eucal}
\usepackage{graphicx}
\usepackage[latin1]{inputenc}
\usepackage[english]{babel}
\usepackage[usenames]{color}
\usepackage[colorlinks=true,linkcolor=blue]{hyperref}
\usepackage{natbib}
\usepackage{nicefrac}
\usepackage{setspace}
\usepackage{url}
\usepackage{lscape}
\usepackage{lineno}
\usepackage{longtable}

\renewcommand{\theta}{\vartheta}
\renewcommand{\phi}{\varphi}
\newcommand{\vect}[1]{\bf{#1}}

\title{Hemispherical Parker waves driven by thermal shear in planetary dynamos}
\author{Wieland Dietrich$^{1,2}$, Dieter Schmitt$^{1,2}$ \& Johannes Wicht$^{2}$ \vspace{0.2cm}
 \\ {\small $ ^1$Department of Applied Mathematics, }
 \\ {\small University of Leeds, Leeds LS2 9JT, United Kingdom}
 \\ {\small $ ^2$Max-Planck-Institut f\"{u}r Sonnensystemforschung,}
 \\ {\small  Max-Planck-Strasse 2, 37191 Katlenburg-Lindau, Germany}}

\begin{document}

\maketitle

\begin{abstract} 
Planetary and stellar magnetic fields are thought to be sustained by
helical motions ($\alpha$-effect) and, if present, differential rotation
($\Omega$-effect). In the Sun, the strong differential rotation in the
tachocline is responsible for an efficient $\Omega$-effect creating a strong
axisymmetric azimuthal magnetic field. This is a prerequisite for Parker dynamo
waves that may be responsible for the solar cycle. In the liquid iron cores of
terrestrial planets, the Coriolis force organizes convection into columns with
a strong helical flow component. These likely dominate magnetic field
generation while the $\Omega$-effect is of secondary importance. Here we use
numerical simulations to show that the planetary dynamo scenario may change
when the heat flux through the outer boundary is higher in one hemisphere than
in the other. A hemispherical dynamo is promoted that is dominated by fierce
thermal wind responsible for a strong $\Omega$-effect. As a consequence Parker
dynamo waves are excited equivalent to those predicted for the Sun. They obey
the same dispersion relation and propagation characteristics. We suggest that
Parker waves may therefore also play a role in planetary dynamos for all
scenarios where zonal flows become an important part of convective motions.
\end{abstract}

\newpage
\section{Introduction}

The dynamo mechanism of stars and planets relies on electromagnetic induction and
requires an electrically conducting medium and a complex fluid flow to maintain
the magnetic fields against Ohmic decay. The liquid iron cores form the dynamo
regions of terrestrial planets and the dominance of Coriolis forces guarantees
sufficiently complicated convective dynamics in these fast rotating bodies.
The flow is organized in convective columns parallel to the rotation axis.
Secondary flows up and down these columns yield a strong helical component
essential for magnetic field generation. Since this helicity associated with
 the local columnar motion dominates the production of the poloidal and the toroidal
magnetic field contributions in many dynamo simulations they are typically
classified as $\alpha^2$ dynamos \cite{Olson1999}. The terminology goes back to
the mean-field dynamo theory where an $\alpha$ stands for the creation of
large-scale field by small-scale motions.

In the Sun, the differential rotation at the tachocline is the main source of
large-scale axisymmetric toroidal magnetic field. Since this is called an
$\Omega$-effect in mean-field dynamo theory, the solar dynamo is classified as
$\alpha \Omega$ \cite{bKrause1980}. A powerful $\Omega$-effect may be the
reason for the oscillatory nature of the solar magnetic field which has not
been observed in planetary dynamos. Strong axisymmetric toroidal fields are a
prerequisite for Parker dynamo waves  \cite{Parker1955} which seem to mimic the
general oscillating behaviour to a large extent.

In the framework of the classical mean-field model of an axisymmetric $\alpha
\Omega$-dynamo \cite{Parker1955}, Parker waves are purely magnetic oscillations
between the poloidal and toroidal mean magnetic field components. Poloidal
field is exclusively created by the $\alpha$-effect as the parametrization of
the nonaxisymmetric helical flows, whereas the azimuthal toroidal field is
induced by the axisymmetric differential rotation or the $\Omega$-effect
\cite{Tobias2007}.

The minimum frequency of these waves is simply the inverse of the magnetic
diffusion time $\tau_\lambda=D^2/\lambda$ \cite{bruediger}. Assuming a
turbulent bulk magnetic diffusivity of $\lambda=10^{8}\, \mathrm{m^2/s}$ for
the solar convective shell of thickness $D=2 \times 10^{8} \, \mathrm{m}$, the
magnetic diffusion time is $\tau_\lambda=12.7\, \mathrm{yr}$ and matches the
time scale of a solar cycle ($22\, \mathrm{yr}$) reasonably well.

Parker waves are typically associated with dynamo models where stress
free flow boundary conditions allow strong zonal winds to arise. These guarantee
a strong $\Omega$-effect and thereby promote Parker waves. Parker-wave like oscillation in combination with a hemispherical magnetic 
field have previously been reported by \cite{Grote2000}. Because of the stress free
flow boundary conditions and the fixed temperature conditions their model also seems
more appropriate for a stellar application.

Here we report that Parker waves also appear in simulations of planetary
dynamos when lateral variations in the heat flux through the outer boundary
lead to fierce zonal winds and thus strong $\Omega$-effects. More specifically,
we consider a model for the ceased ancient Martian dynamo 
which left its trace in form of a remanent crustal magnetization. The
magnetization shows a hemispherical distribution with much higher values in the
southern than in the northern hemisphere \cite{Acuna1999}. Recent studies aim
at explaining this pattern with the fact that the dynamo itself produced a
hemispherical field \cite{Stanley2008,Dietrich2013}. Such dynamos are
promoted when the heat flux through the southern core-mantle boundary is
significantly larger than that through the northern. In terrestrial planets,
the core-mantle-boundary heat flux is determined by the thermal mantle
structure which develops on much slower time scales than core convection.
A major impact in
the northern hemisphere or a large-scale mantle convection pattern are two
alternative scenarios causing a north/south heat flux asymmetry \cite{Stanley2008}.

Since the southern hemisphere of the core is cooled more efficiently than its
northern counterpart, a latitudinal temperature gradient develops that drives
fierce thermal winds.  When the heat flux asymmetry is large enough these winds
clearly dominate the flow. The dynamo then changes from an $\alpha^2$ to an
$\alpha\Omega$ type and shows wave-like behaviour reminiscent of Parker waves
\cite{Dietrich2013}. Here we analyse a suite of hemispherical dynamo
simulations to determine when these waves arise. By showing that the frequency
of the waves follows the theoretical dispersion relation based on mean-field
theory we clearly establish their Parker wave nature.

\section{Model}

The full MHD dynamo problem is defined by the equations for conservation of
momentum, for the evolution of the superadiabatic temperature
perturbation $T$ and for the evolution of the magnetic field $\vect{B}$. For a
rotating spherical shell and assuming the Boussinesq-approximation these
equations read
\begin{align}
& \frac{E}{Pm} \left( \frac{\partial \vect{u}}{\partial t} + \vect{u} \cdot
\vect{\nabla} \vect{u}\right) - E\vect{\nabla}^2 \vect{u} = -\vect{\nabla} \Pi \label{nseq} \\
& \hspace*{1cm} - 2 \hat{\vect{z}} \times \vect{u} + Ra_q^\star Pm \frac {\vect{r}}{r_{o}} T
+ (\vect{\nabla} \times \vect{B} ) \times \vect{B} \, \nonumber \\
& \frac {\partial T}{\partial t} + \vect{u} \cdot \vect{\nabla} T = \frac {Pm} {Pr} \nabla^2 T
+ \epsilon \label{heat} \\
& \frac {\partial \vect{B}}{\partial t} - \vect{\nabla} \times \left( \vect{u} \times \vect{B} \right)
= \vect{\nabla}^2 \vect{B} \label{ind_eq}
\ .
\end{align}
Here $\vect{u}$ is the velocity, $\Pi$ is the non-hydrostatic pressure, $\epsilon$ a homogeneous heat
source density that models the cooling of the planet, and $\hat{\vect{z}}$ the
unit vector parallel to the axis of rotation. Above equations have been made
dimensionless using the magnetic diffusion time $D^2/\lambda$ as a time scale
and the shell thickness $D=r_o-r_i$ as a length scale, where $r_o$ and $r_i$
are the outer and inner shell radii, respectively, and $\lambda$ is the
magnetic diffusivity. The magnetic scale is $(\mu\lambda\rho\Omega)^{1/2}$ with
$\mu$ the magnetic permeability, $\rho$ the mean density, and $\Omega$ the
rotation rate.

The dimensionless control parameters are the Ekman number $E=\nu / \Omega D^2$,
a measure of the ratio of viscous to Coriolis force, the modified flux based
Rayleigh number $Ra_q^\star=Ra_q E / Pr=\beta q_o g_o D^2 / \rho c_p \nu \kappa \Omega$,
the hydrodynamic Prandtl $Pr=\nu / \kappa$ and magnetic Prandtl number $Pm=\nu
/ \lambda$. Here, $\kappa$ is the thermal diffusivity, $\nu$ the kinematic
viscosity, $g_o$ the gravity at the outer boundary, $\beta$ the thermal
expansivity, $\rho$ the density, $c_p$ the specific heat and $q_0$ the 
superadiabatic outer boundary heat flux.

The choice of rigid flow boundaries and an electrically insulating mantle 
and inner core is motivated by the application to terrestrial 
planets \cite{Dietrich2013}. To more specifically model the ancient Martian 
dynamo where likely no inner core was present we drive convection by homogeneously 
distributed heat sources $\epsilon$ which is equivalent to secular cooling. For 
numerical reasons, we kept an inner core with an aspect ratio of $r_i/r_o=0.35$ 
in our simulations but set the inner boundary heat flux to zero to minimize its impact. 
The total heating $V\epsilon$ is balanced by the total heat loss through the outer
boundary $4\pi r_o^2 q_0$, where $V$ is the volume of the outer core and $q_0$
is the mean heat flux density through the outer boundary. Further to modify the
heat escape out of the spherical shell we assume a cosine variation of the
outer boundary heat flux like
\begin{equation}
q=q_0 ( 1 - q^\star \cos{}\theta )\;\;,
\end{equation}
where $\theta$ is the colatitude and $q^\star$ the relative variation
amplitude. Thus setup was first proposed by \cite{Stanley2008}.

For the bulk of the numerical results the nondimensional parameters are fixed
to $E=10^{-4}, Ra_q=4.1 \times 10^{7}, Pm=2, Pr=1$ and we gradually increase
$q^\star$ up to $3$. All simulations were performed with the MagIC3 code
\cite{Wicht2002} using a resolution $N_r \times N_\phi \times N_\theta$ 
of maximal $65 \times 320 \times 160$. After equilibration each model ran for at least
one (typically more) magnetic diffusion time. We avoided the usage of azimuthal symmetries
or hyperdiffusion to speed up the simulations.

\subsection{Mean-field theory}

Mean-field theory focusses on solving the large-scale induction equation by
parameterizing the effects of small-scale magnetic field and flow. Using
appropriate averaging techniques, magnetic field and flow are both separated
into a mean or large-scale contribution, denoted by overbars, and fluctuating
or small-scale contribution, denoted by primes:
\begin{align}
\vect{u}=\overline{\vect{u}} + \vect{u}' , \, \vect{B}=\overline{\vect{B}} + \vect{B}'
\end{align}

\begin{figure}
\centering
\includegraphics[width=0.75\textwidth]{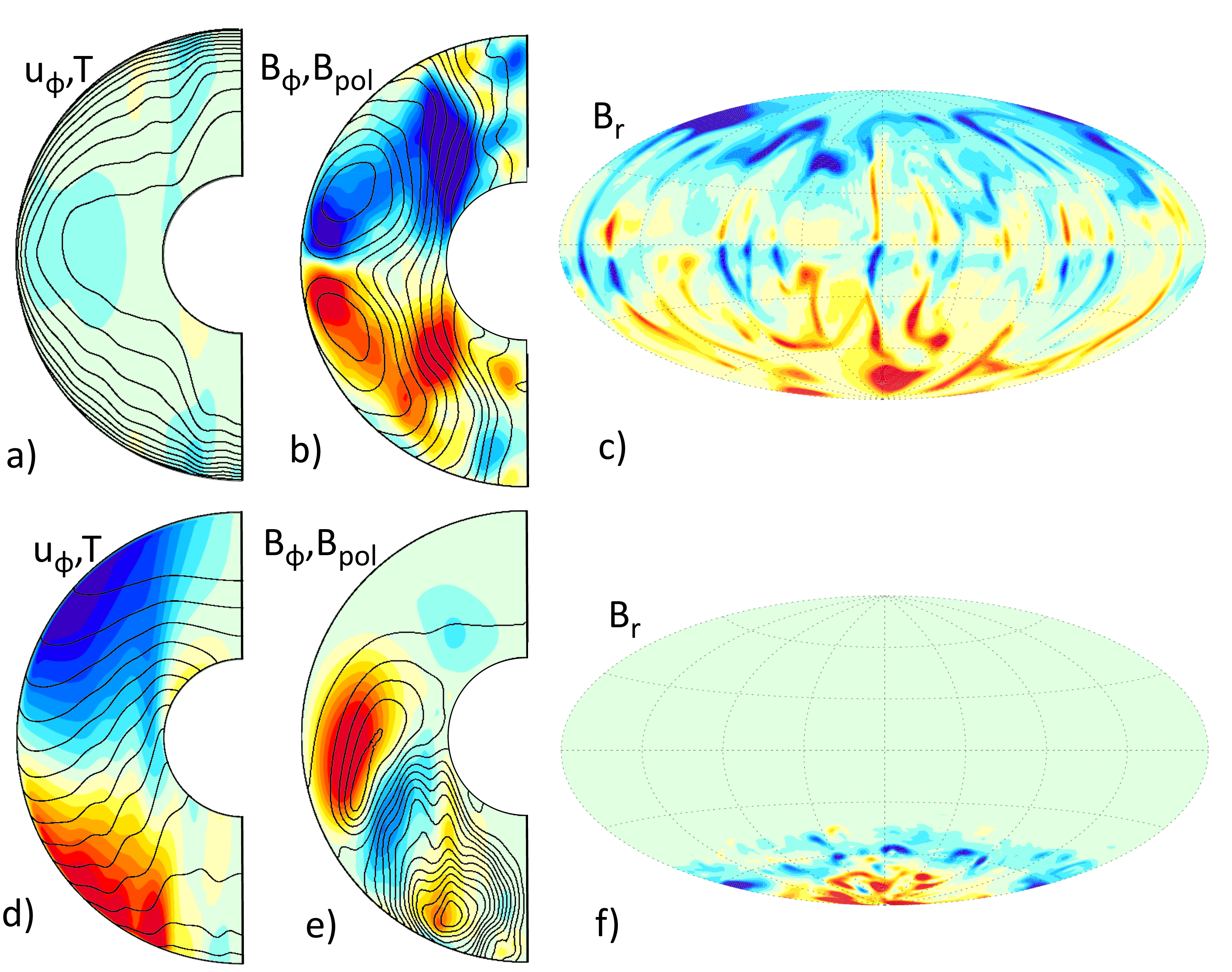}
\caption{Comparison of the homogeneous reference case (upper row) and a
hemispherically modified dynamo (lower row). The left plots (a and d) show
the colour coded zonal flow with the axisymmetric temperature as contours,
the middle plot characterizes the azimuthal field in colour and the poloidal
field lines as contours (b and e). The right plots display the radial field
at the outer boundary (c and f). All plots are snapshots.}
\label{zonalshort}
\end{figure}

For our purpose we use azimuthal averages so that $\overline{\vect{u}}$ and
$\overline{\vect{B}}$ are the axisymmetric and $\vect{u}'$ and $\vect{B}'$ are
the non-axisymmetric contributions. In the induction equation for the mean
field $\overline{\vect{B}}$, we reduce the action of $\overline{\vect{u}}$ to
the $\Omega$-effect since other contributions are likely much smaller. The
action of the fluctuating flow $\vect{u}'$ is parameterized by the
$\alpha$-effect, assuming homogeneous turbulence \cite{bKrause1980}. Following
recent studies, such as \cite{Busse2006} and \cite{Schrinner2011}, we
investigate axisymmetric $\alpha\Omega$ and $\alpha^2\Omega$-dynamos
responsible for creating an axisymmetric mean field $\overline{\vect{B}}=B
\vect{e}_\phi + \vect{\nabla} \times A \vect{e}_\phi$, with toroidal vector
potential $A$, representing the poloidal field, and toroidal field $B$:
\begin{align}
\frac{\partial A}{\partial t} &= \alpha B +  \Delta A \\
\frac{\partial B}{\partial t} &= -\alpha \Delta A + \frac{\partial \overline{u}}
{\partial x_2} \frac{\partial A}{\partial x_3}  + \Delta B \ .
\label{eqmeanfieldfull}
\end{align}
For remaining consistent with previous formulations \cite{Schrinner2011}, we
use cartesian coordinates $x_1,x_2,x_3$ corresponding to the spherical
$\phi,\theta,r$, respectively. As a solution for this system of equations,
Parker \cite{Parker1955} suggested propagating plane dynamo waves obeying the
following dispersion relations for the frequency $\nu$ in the $\alpha\Omega$
and $\alpha^2\Omega$ limit:
\begin{align}
\nu_{\alpha \Omega} &= \sqrt{\frac{\alpha}{2}} \sqrt{k_3 \frac{\partial
\overline{u}}{\partial x_2}} \ , \label{dispaw} \\
\nu_{\alpha^2 \Omega} &= \sqrt{\frac{\alpha}{2}} \sqrt{ \sqrt{ \left(\alpha k^2\right)^2
+ \left(k_3 \frac{\partial \overline{u}}{\partial x_2}\right)^2 } - \alpha k^2} \ , \label{dispa2w}
\end{align}
The total wave number $k$ with $k^2=k_2^2+k_3^2$ has the two contributions
$k_2$ and $k_3$ in the $\theta$- and $r$-direction, respectively.

\begin{figure}
\centering
\includegraphics[width=0.75\textwidth]{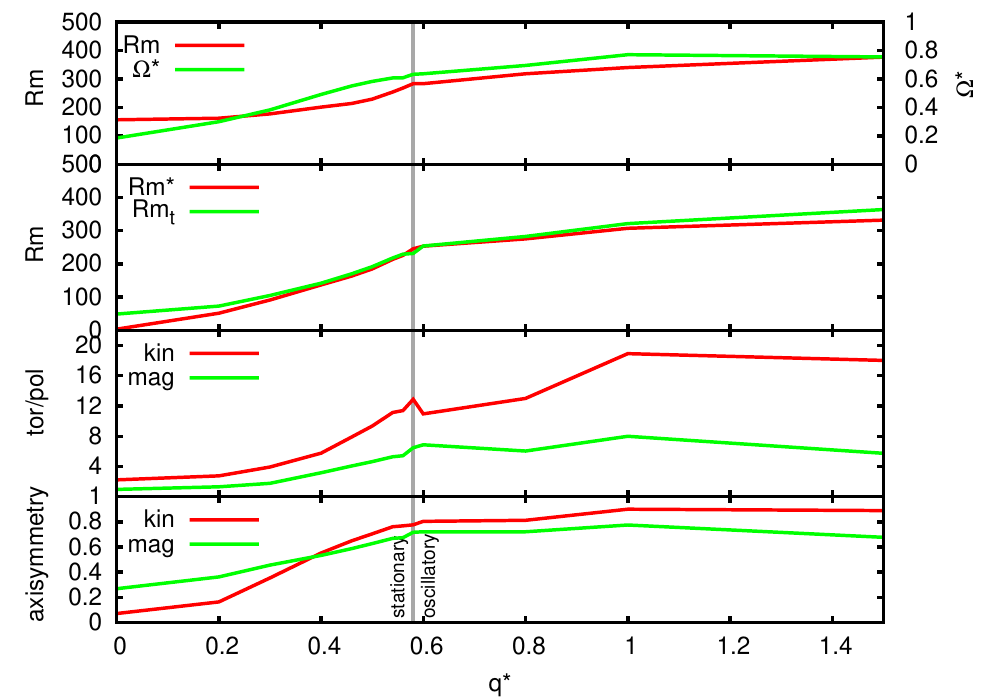}
\caption{First panel: magnetic Reynolds number and the relative amount of
toroidal field induced by the axisymmetric $\Omega$-effect as a function of the
heat flux anomaly amplitude $q^\star$. Second panel: magnetic Reynolds number
based only on the axisymmetric and equatorially antisymmetric zonal flows
$Rm^\star$ and the strength of thermal wind $Rm_t$ according to eq.
\ref{eqrmt}. Third panel: ratio of toroidal to poloidal kinetic and magnetic
energy. Bottom panel: relative axisymmetry in the kinetic and magnetic energy.
The gray vertical line at $q^\star \approx 0.6$ denotes the onset of the
oscillations.}
\label{mp_hemis}
\end{figure}

\section{Results}

Figure \ref{zonalshort} illustrates the effect of the latitudinal heat flux
variation on the convection and induction mechanism.  For a homogeneous heat
flux the temperature and the weak zonal flow are symmetric with respect to the
equator (fig. \ref{zonalshort} a), whereas the magnetic field is antisymmetric
(fig. \ref{zonalshort} b and c). If a sufficiently strong heat flux anomaly is
applied, here $q^\star=1.0$, the temperature shows a strong gradient in
latitudinal direction that penetrates deep into the shell (fig.
\ref{zonalshort} d). Northward directed flows tend to equilibrate the
latitudinal temperature anomaly and are diverted by the Coriolis force into the
azimuthal direction. This effect is known as thermal wind and relates
latitudinal gradients in the axisymmetric temperature $\overline{T}$ with zonal
flow $\overline{u}_\phi$ gradients parallel to the axis of rotation
\begin{align}
\frac{\partial \overline{u}_\phi}{\partial z} = Ra_q^\star  Pm 
\frac{1}{2 r_o} \frac{\partial \overline{T}}{\partial \theta} \ . \label{eqthermalwind}
\end{align}
Toroidal field is mainly produced by an $\Omega$-effect associated with the shear
between the retrograde zonal flow in the northern hemisphere and the prograde
zonal flow in the southern hemisphere (fig. \ref{zonalshort} e). Convective
columns are largely missing and the small-scale convective flow is dominated by
plume-like up- and downwellings in the southern hemisphere. Since these are the
major source for helical motions, the $\alpha$-effect and thus poloidal field
production is concentrated in the southern hemisphere (fig. \ref{zonalshort} e
and f).

\begin{figure}
\centering
\includegraphics[width=0.5\textwidth]{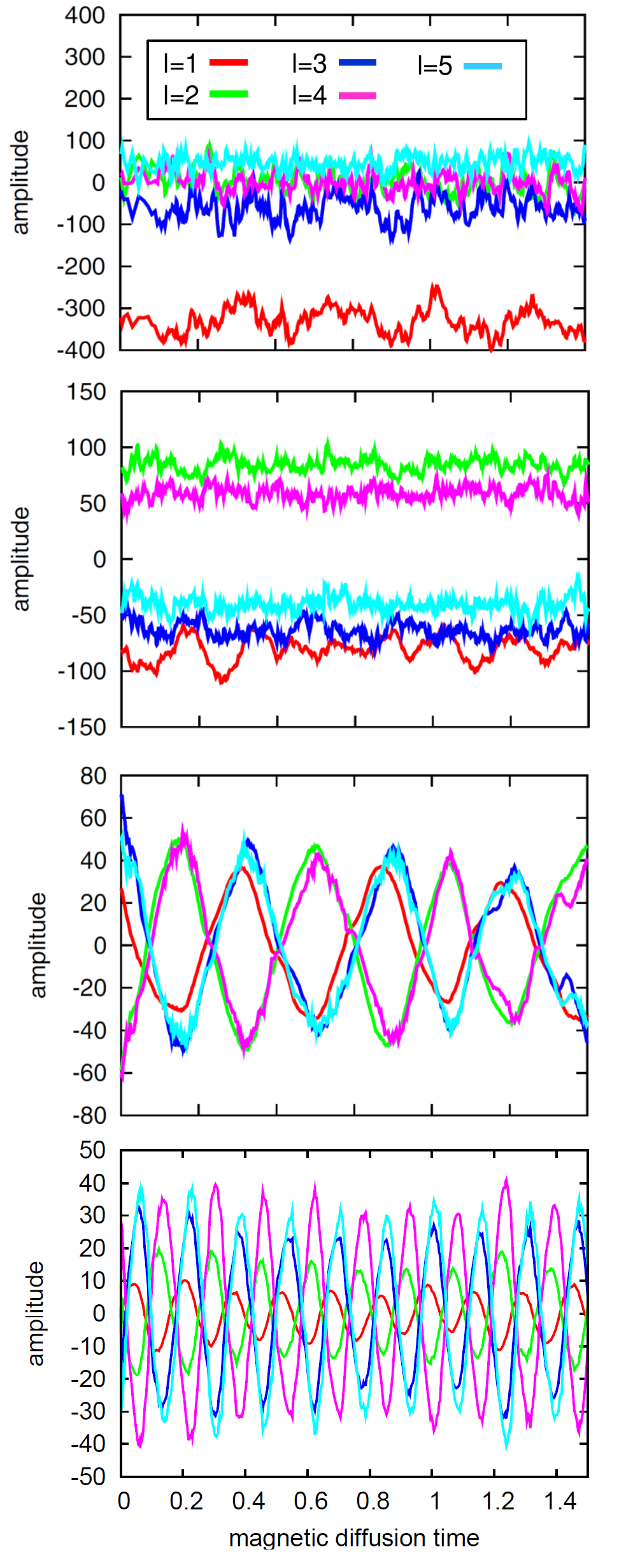}
\caption{Time evolution of the Gauss coefficients for $q^\star=0,0.5,0.6,1$
(from top to bottom). Colour coded are the first five axisymmetric coefficients
of degree $l=1,2,3,4,5$ in red, green, dark blue, pink and light blue.}
\label{gaussevo}
\end{figure}

\begin{figure}
\centering
\includegraphics[width=0.75\textwidth]{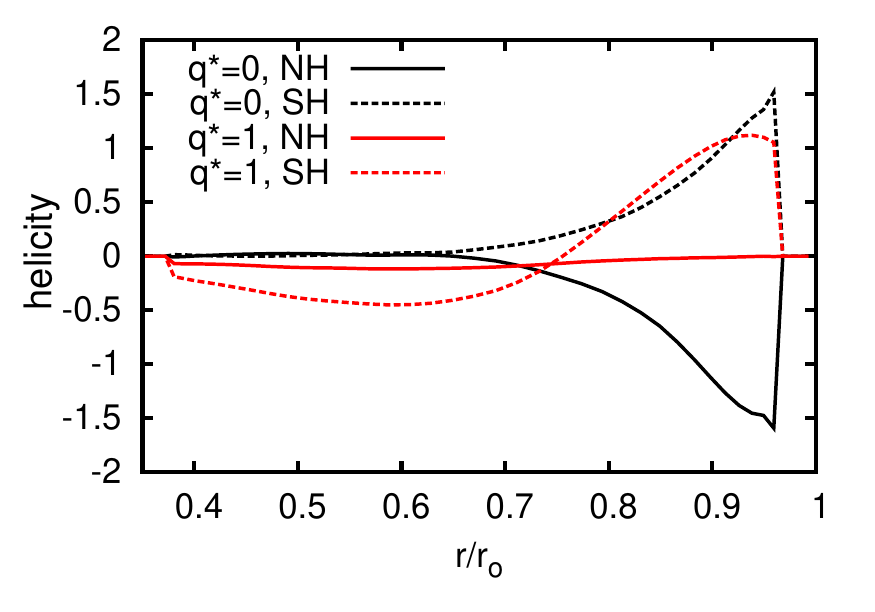}
\caption{Radial profiles of the horizontally averaged kinetic helicity for the
southern (SH) and northern hemisphere (NH).}
\label{helrad}
\end{figure}

Fig. \ref{mp_hemis} demonstrates how the flow and the magnetic field becomes
predominantly axisymmetric and toroidal when the amplitude $q^\star$ increases.
Since the axisymmetric toroidal flow contribution is identical to the
axisymmetric azimuthal (or zonal) flow this illustrates the dominant role of
thermal winds at larger $q^\star$ values. In the scaling used here, the
magnetic Reynolds number $Rm= u  D/ \lambda$ corresponds to the velocity. In
fig. \ref{mp_hemis}, second panel we test the influence of the thermal wind by
comparing the flow amplitude based on the equatorially antisymmetric and
axisymmetric flow $Rm^\star$ with the amplitude of the thermal wind $Rm_t$
controlled by
\begin{align}
Rm_t = Ra_q^\star Pm \frac{l_z}{2 r_o} \left\langle\frac{\partial \overline{T}}{\partial \theta}
\right\rangle_{\mathrm{rms}} \ .
\label{eqrmt}
\end{align}
Here, $l_z$ is the typical length scale for zonal flow variations in the
direction of the rotation axis and is taken to be half the height of the
spherical shell at mid-depth ($l_z\approx 1.18$). The good agreement shows that
the equatorially antisymmetric zonal flow originates from thermal wind alone.
In the regime $q^\star\ge 0.6$ both the thermal wind magnetic Reynolds number
estimate $Rm_t$ and $Rm^\star$ approaches the true magnetic Reynolds number
$Rm$ as is shown in fig. \ref{mp_hemis}. This demonstrates the thermal wind
origin of the fierce zonal flows.

We further quantify the relative importance of the $\Omega$-effect in toroidal
field production by :
\begin{align}
\Omega^\star= \frac { \langle \vect{B} \cdot \vect{\nabla}  \overline{u}_\phi
\rangle^{\mathrm{tor}}_{\mathrm{rms}}}  { \langle \vect{B} \cdot \vect{\nabla}
\vect{u} \rangle_{\mathrm{rms}}^{\mathrm{tor}}}  \ . \label{defomega}
\end{align}
Fig. \ref{mp_hemis} illustrates that $\Omega^\star$ indeed increases beyond
$0.5$ once $q^\star\ge 0.6$. For a homogeneous heat flux ($q^\star=0$) the
$\Omega$-effect is negligible. The latitudinal heat flux variation thus clearly
changes the dynamo mechanism from an $\alpha^2$ to an $\alpha^2 \Omega$ or
$\alpha \Omega$ type. The effective $\Omega$-effect has the consequence that
the magnetic field is dominated by the axisymmetric toroidal magnetic field as
is shown in the lower panels of fig. \ref{mp_hemis}. Strong Lorentz forces
associated to this field component nearly completely suppress the convective
columns at larger $q^\star$ values \cite{Dietrich2013, Stanley2008}.

\begin{figure}
\centering
\includegraphics[width=0.75\textwidth]{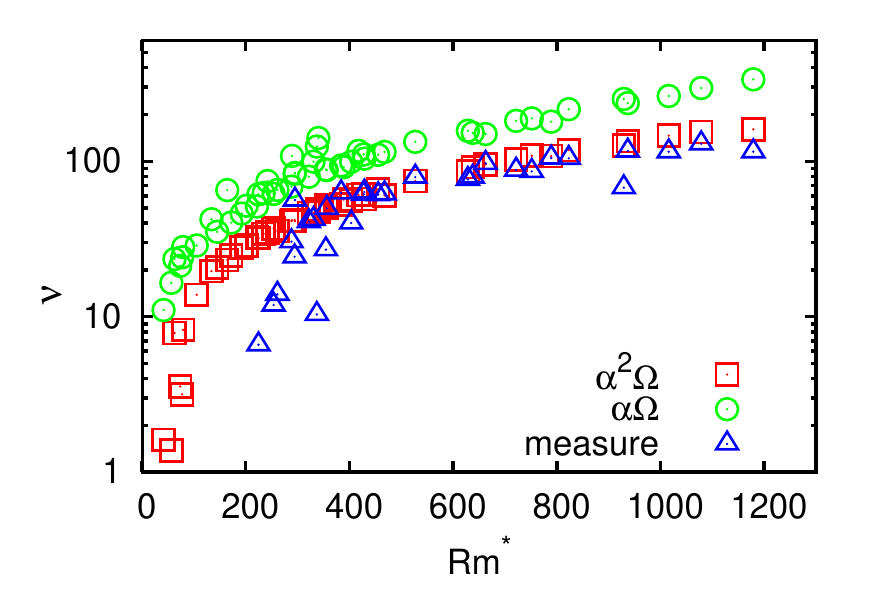}
\caption{Frequencies measured from the evolution of the Gauss coefficients
(blue triangles), the frequencies calculated by the $\alpha \Omega$
(eq. \ref{dispaw}, green circles) and the $\alpha^2 \Omega$
(eq. \ref{dispa2w}, red squares) dispersion relation as function of the zonal
flow amplitude given by $Rm^\star$. }
\label{parkerboth}
\end{figure}

The onset of the dynamo wave is roughly found where the $\Omega$-effect starts
to play a dominant role at $q^\star\approx0.6$. Fig. \ref{gaussevo} shows the
time evolution of the leading axisymmetric Gauss coefficients describing the
magnetic field at the top of the dynamo region. In the reference case
($q^\star=0$) the dipole contribution clearly dominates. When the perturbation
is increased to $q^\star=0.5$, the even and odd modes reach similar amplitudes
but have opposite sign. This reflects a hemispherical magnetic field
concentrated in the southern hemisphere (2nd panel). The time dependence is
still ruled by the turbulent convective flow dynamics as in the reference case.
A further increase of the heat flux variation to $q^\star=0.6$ finally promotes
the onset of the dynamo wave. Even and odd modes oscillate periodically in
antiphase with nearly similar amplitudes. If $q^\star$ is further increased,
both the thermal wind shearing (see $Rm^\star$ in fig. \ref{mp_hemis}) and the
frequency of the oscillation grow. This is the expected behaviour of a
mean-field Parker wave \cite{Schrinner2007}, as shown in the dispersion
relation (eqs. \ref{dispaw} and \ref{dispa2w}). In a numerical test, we
switched off the Lorentz force in our simulation. The oscillation persisted as
a purely kinematic effect which is a key feature of Parker waves.

To more qualitatively test whether the os<cillation obeys the Parker wave
dispersion relation (eq. \ref{dispaw} and \ref{dispa2w}) we have to estimate
the wave numbers, the $\Omega$-effect or latitudinal shear, and the
$\alpha$-effect via the helicity \cite{Tobias2007}. For the large-scale wave
behaviour we simply approximate the wave numbers $k_2, k_3$ with $2\pi$-th of
the inverse characteristic length scales $l_i$. We choose $l_2\approx \pi$ as
half the circumference along latitude at mid-depth, and $l_3 = 1$ as the shell
thickness. The latitudinal shear is estimated with $\partial \overline{u} /
\partial x_2 \approx Rm^\star / \pi $. The $\alpha$-term is approximated via
$\alpha= - H \,\tau /3 $ where $\tau$ is the correlation time and $H$ is the
kinetic helicity \cite{bKrause1980}.

We further assume that $\tau$ is the mean flow overturn time $\ell/Rm^\prime$
where $Rm^\prime$ quantifies the mean amplitude of the non-axisymmetric flow
and $\ell$ is its typical length scale estimated from the spectra of the
poloidal flow.

\begin{figure}
\centering
\includegraphics[width=0.75\textwidth]{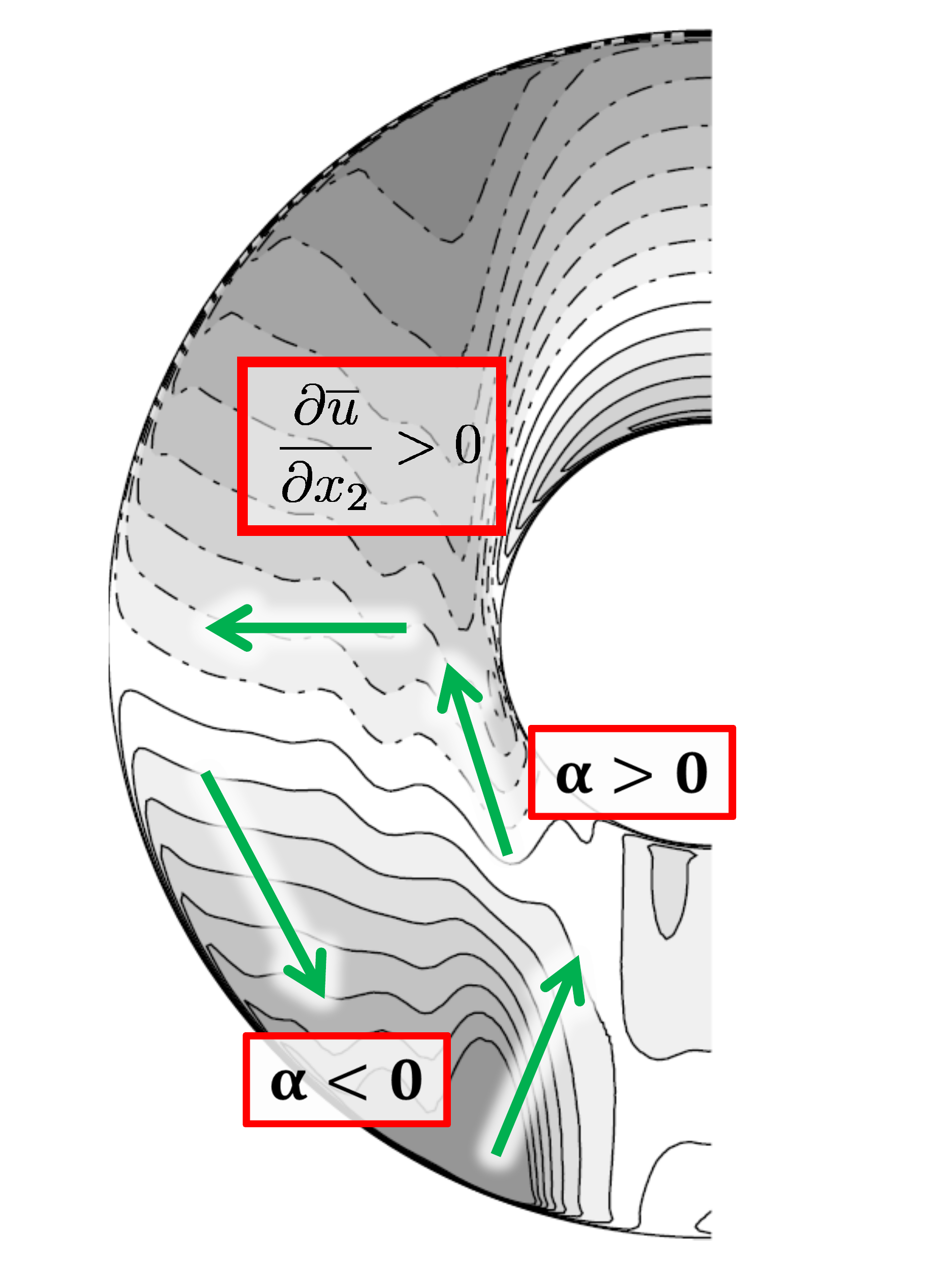}
\caption{Angular velocity in gray scale, signs of $\alpha$ and $\Omega$-effect
(red) and theoretical propagation direction of $\alpha \Omega$ waves (green
arrows).} \label{parker_zonal}
\end{figure}

Estimating the helicity is more involved. Fig. \ref{helrad} shows radial
profiles of the rms helicity in the northern and southern hemispheres for the
homogeneous reference case and the hemispherical $q^\star=1.0$ case. The
helicity $H=\vect{u}' \cdot \vect{\nabla} \times \vect{u}'$ is averaged either
over the southern (dashed lines) or the northern hemisphere (solid). 
Note that only the non-axisymmetric flow components are used here. In
the reference case ($q^\star=0$, black) the helicity is negative in the
northern, positive in the southern hemisphere and antisymmetric with respect to
the equator. Such a configuration is essential for creating dipole dominated
magnetic fields. In the hemispherical solution, the northern hemisphere is
devoid of helicity (red solid), whereas in the southern hemisphere (red dashed)
the sign changes from negative in the inner part to positive in the outer part
of the dynamo shell. Since the dynamo wave is concentrated to the southern
hemisphere we use only the rms helicity in this hemisphere to estimate
$\alpha$.

In fig. \ref{parkerboth} we compare the predicted frequencies $\nu_{\alpha
\Omega}$ and  $\nu_{\alpha^2 \Omega}$ according to eqs. \ref{dispaw} and
\ref{dispa2w} with the measured frequency for $51$ simulations at different
parameters. They cover Ekman numbers from $E= 3 \times 10^{-4}$ to $E=3 \times
10^{-5}$, magnetic Prandtl numbers from $Pm=1$ to $Pm=5$, and a broad range of
Rayleigh numbers and $q^\star$ values. The zonal flow magnetic Reynolds number
$Rm^\star$ used for the $x$-axis of fig. \ref{parkerboth} mainly depends on
$q^\star$. Since the dynamo waves only set in at larger $q^\star$ values there
are no measured frequencies at low $Rm^\star$. Also we expect an improved fit
to the prediction at larger $Rm^\star$ values where axisymmetric toroidal field
and flow clearly dominate. Considering the various simplifications and
approximations involved the agreement between measured and predicted
frequencies seems convincing enough to establish the Parker wave nature of the
oscillations. The $\alpha^2 \Omega$ frequencies $\nu_{\alpha^2 \Omega}$ are
always smaller than the $\nu_{\alpha \Omega}$, since the second $\alpha$-effect
reduces the frequency. However, the data scatter seem too large to distinguish
whether the mechanism is of the $\alpha \Omega$- or the  $\alpha^2
\Omega$-type.

\begin{figure*}
\begin{center}
\includegraphics[width=0.75\textwidth]{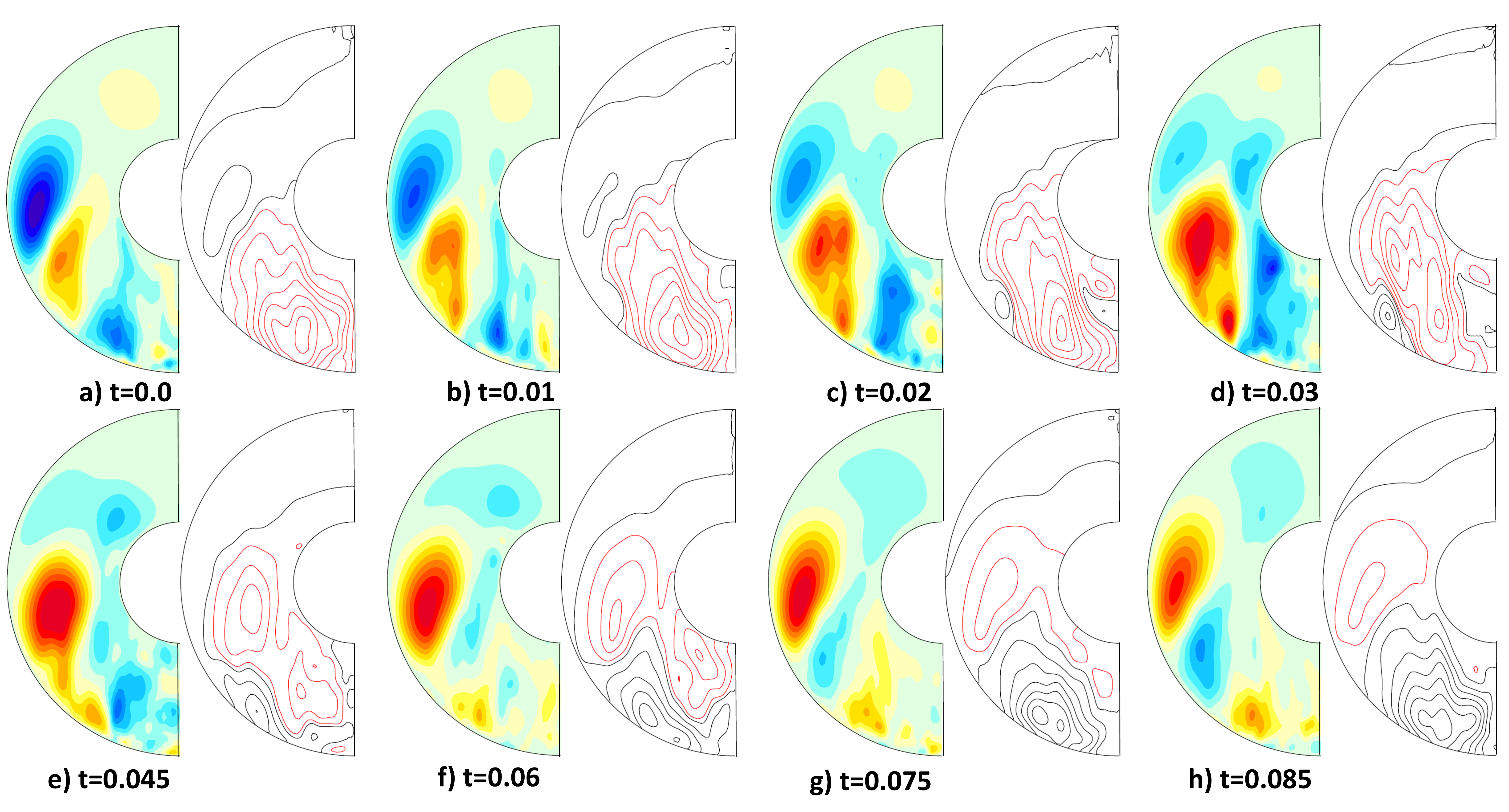}
\caption{Temporal evolution of the axisymmetric toroidal field (left) and
poloidal field lines (right) during half a wave cycle. Note, that the time
difference between two shown snapshots changes non-equidistantly.}
\label{wave_evo}
\end{center}
\end{figure*}

The propagation behaviour is another key spotting feature for Parker waves
\cite{Yoshimura1976}. In general, a Parker wave travels along lines of constant
shear and its direction depends on the sign of the product of the $\alpha$- and
$\Omega$-effect \cite{Tobias2007}. Since the northern zonal flow cell is
progressing westward (slower than average) and the southern eastward (faster
than average) the $\Omega$-effect $\partial \overline{u} / \partial x_2$ is
positive everywhere (see fig. \ref{parker_zonal}). The helicity changes its
sign at some depth in the shell for strongly hemispherical solutions (see fig.
\ref{helrad}). Fig.~\ref{parker_zonal} illustrates the resulting propagation
path expected for a Parker wave in the hemispherical cases. This is consistent
with the observed propagation of the dynamo wave shown in fig.~\ref{wave_evo}.
Suppose the wave cycle starts at lower southern latitudes close to the outer
boundary (small red patch, fig. \ref{wave_evo}, first plot) with a poleward and
inward migration of the field. Due to the strong $\alpha$-effect the field is
amplified and propagates further northwards, replacing the inverse polarity. At
the equator the wave migrates along lines of constant shear (radially outward
here).

\section{Discussion}

We demonstrated that the oscillating solution found in strongly hemispherical
dynamo solutions promoted by a north/south heat flux asymmetry through the
outer boundary are likely Parker dynamo waves. The heat flux pattern leaves the
northern hemisphere hotter than the southern and the associated temperature
gradient drives fierce zonal thermal winds. These in turn lead to a strong
$\Omega$-effect, a prerequisite for Parker waves.

In the Martian context this is bad news because the relatively fast
oscillations on the order of a few thousand years are not compatible with the
high crustal magnetization amplitude \cite{Dietrich2013}. Since crustal
magnetization is acquired over several million years the effective
magnetization as observed from a space craft would be very weak.

Lateral heat flux variations with $q^\star\approx 1$ seem generally possible
for terrestrial planets, at least as long as the mantle is actively convecting.
The large-scale cosine-like variation explored here, however, is geared to
explain the Martian hemisphericity. For Earth, the heat flux pattern is
dominated by a spherical harmonic of degree and order two. This was likely
different in the past, but a degree one pattern nevertheless seems more likely
for Mars while higher harmonics are expected for Earth \cite{Roberts2006}.
Another property essential for promoting strong thermal winds is the dynamo
heating mode. The effect of an outer boundary heat flux variation is much
larger for a dynamo without an inner core that is exclusively driven by secular
cooling \cite{Hori2012}. Whether a combination of a more complex heat flux
pattern and secular cooling would also lead to strong thermal winds and
possibly Parker waves has not been explored in any detail so far. This may have
been the case for the geodynamo before the onset of inner core growth which may
have happened only one Gyr ago.

Recent simulations geared to model the dynamo in the gas giants also show
oscillatory behaviour reminiscent of Parker waves \cite{Gastine2012}. Here the
stress free outer boundaries allow much stronger zonal winds to develop than in
typical simulations for terrestrial dynamos. This can also lead to a strong
$\Omega$-effect and associated Parker wave behaviour. Parker waves are thus not
only interesting to explain the solar cycle but also have to be considered in
the planetary context.

\bibliographystyle{plainnat}
\bibliography{parkerpaper}

\end{document}